\documentclass[pre,showpacs,twocolumn]{revtex4}
\usepackage{graphicx,amssymb,color}
\usepackage[dvipsnames]{xcolor}
\usepackage[colorlinks=true,linkcolor=red,citecolor=red]{hyperref}
\usepackage{siunitx}
\usepackage{enumerate}

\begin{document}
\title{Generation of intermittent gravito-capillary waves via parametric forcing}
\author{Gustavo Castillo}
\affiliation{Instituto de Ciencias de la Ingenier\'ia, Universidad de O'Higgins, Av. Libertador Bernardo O'Higgins 611, Rancagua, Chile}
\email[Corresponding author: ]{gustavo.castillo@uoh.cl} 
\author{Claudio Falc\'on}
\affiliation{Departamento de F\'isica, Facultad de Ciencias F\'isicas y Matem\'aticas, Universidad de Chile, Avenida Blanco Encalada 2008, Santiago, Chile}
\date{\today}

\begin{abstract}

We report on the generation of an intermittent wave field driven by a horizontally moving wave maker interacting with Faraday waves. The spectrum of the local gravito-capillary surface wave fluctuations displays a power-law in
frequency for a wide range of forcing parameters. We compute the probability density function of the local surface height increments, which show that they change strongly across time scales. The structure functions of these increments are shown to display power-laws as a function of the time lag, with exponents that are nonlinear functions of the order of the structure function. We argue that the origin of this scale-invariant intermittent spectrum is the Faraday wave pattern breakup due to its advection by the propagating gravity waves. Finally, some interpretations are proposed to explain the appearance of this intermittent spectrum.

\end{abstract}
\maketitle

\section{Introduction}

The nature, generation, properties and evolution of turbulent fields in extended out-of-equilibrium systems is the subject of continuous research due to its implications in transport of conserved quantities across scales in fluid dynamics, condensed matter, plasma and astrophysics. In the case of wave systems, this research has been strongly driven by the development of the tools of Wave Turbulence (WT)~\cite{KolmogorovWTBook}. WT deals with the long time out-of-equilibrium statistical properties of dispersive waves in weakly nonlinear interaction, in which injection and dissipative scales are clearly separated, and energy is transferred without loss among scales (in the so-called inertial window). In the WT framework, waves are thus solely described by their scale-invariant stationary spectrum of wave amplitudes  $S_{\eta}(f)=\langle|\eta_{f}|^{2}\rangle\propto f^{-\nu}$, where $f$ is the frequency and $\nu$ the Kolmogorov-Zakharov (KZ) exponent. This spectrum has been theoretically predicted, numerically computed and experimentally observed in gravity~\cite{Hasselmann1962,ZakharovDiachenko2004,Toba1973,FauveFalcon2007} and capillary~\cite{ZakharovFilonenko1967,PushkarevZakharov1995,Putterman1996} surface waves in fluids, bending waves in elastic plates~\cite{DuringRica2006,Mordant2008}, and nonlinear optics~\cite{Dyachenko1992,Bortolozzo2009}. Applications of WT to a larger number of other fields can be found in~\cite{KolmogorovWTBook}. 

It may seem that WT is a robust and complete theory that can be applied on any weakly dispersive nonlinear wave system, but this is not the case as it is almost never valid over all length scales~\cite{Biven2001}: there is a breakdown of WT at very large or very low wave numbers. When the breakdown occurs, a new type of spectrum develops, as well as new properties of the turbulent fluctuations~\cite{During2009}, which are different from the WT prediction. This type of breakdown spectrum, which is named after Phillips seminal work on the breakup of gravity surface waves by wind forcing~\cite{Phillips1958}, has been observed {\it in-situ} on the sea surface, in very large tanks~\cite{Mitsuyasu1974,Romero2012} and in laboratory experiments~\cite{FauveFalcon2007,NazarenkoLukaschuk2010}. It has been generalized to other situations where WT breaks down~\cite{NewellZakharov2008} such as the generation of rough sea foam in the ocean~\cite{NewellZakharov1992} and the {\it d-}cone spectra in vibrating elastic plastes~\cite{Miquel2013}. It has also motivated new ideas on the spectra generated by singularities~\cite{Kuznetsov2004} to describe the properties of fluctuating wave fields.

WT can also break down when the locality of interactions between wave vectors is forbidden. In the WT framework, a conserved quantity is transported by local interactions in wave vector space from large scales to small scales. However, this is not always achieved in real systems. Kelvin waves in superfluids are known to interact non locally through the Biot-Savart equation~\cite{Laurie2010,Lvov2010}. In strongly magnetized plasma, it has been proven that its dynamics is dominated by the nonlinear, nonlocal interaction between the large scale condensate and small scales~\cite{Balk1990,Nazarenko1991,Connaughton2011}. Wave turbulence in quantum field theory (QCD) can also be described by nonlocal interactions in momentum space \cite{Mehtar2017}.

Here, we present an experimental study of intermittency in which spatially extended standing waves of wave vector $k_o$ (Faraday waves) interact nonlinearly with a random set of long waves generating a scale invariant wave spectrum which differs from the KZ predictions for gravity wave turbulence or capillary wave turbulence. The local measurement of the wave amplitude field displays singular events which can be related to singularities generated due to the advection Faraday waves by random surface gravito-capillary waves. In this configuration, the local temporal fluctuations of the surface wave amplitude $\eta(t)$ display a power-law spectrum $S_{\eta}(f)=\langle|\eta_{f}|^{2}\rangle$ ({\it i.e.}, the  Fourier transform of the autocorrelation function of $\eta(t)$) over more than half a decade for a wide range of forcing parameters. The observed spectrum $S_{\eta}(f)\propto f^{-5}$ differs from the one predicted by the WT theory $S_{\eta}(f)\propto f^{-4}$. This wave field displaying this spectrum is intermittent, which is shown by computing the probability density function (PDF) of the local vertical height increments and their corresponding structure functions. The PDFs change strongly across time scales, increasing their flatness. The structure functions of these increments display power-laws as a function of the time lag, with exponents that are not linear with the order of the structure function. We argue that the origin of this scale-invariant intermittent spectrum is the Faraday wave pattern breakup due to its advection by the propagating gravity waves. 

\section{Experimental setup}

\begin{figure}[ht!]
\centering
\includegraphics[width=\columnwidth]{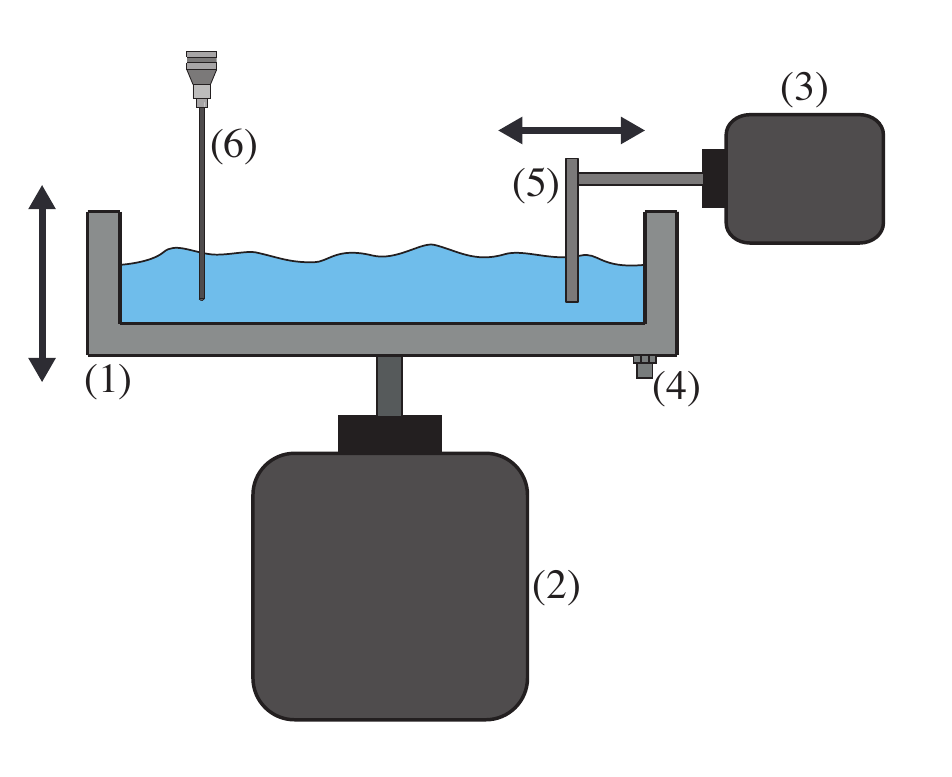}
\caption{(color online) Schematic of the experimental setup. (1) Square cell, (2) Vertical shaker, (3) Horizontal shaker, (4) Accelerometer, (5) Wave-maker, (6) Capacitive wire gauge.}
\label{FigSetup}
\end{figure}

The experimental setup is depicted in Fig. \ref{FigSetup}. A plexiglass square cell (lateral dimensions $L_{x}=L_{y}=\SI{10}{\centi\meter}$ and height $L_{z}=\SI{4.5}{\centi\meter}$) is filled with distilled water up to a height $h=\SI{3}{\centi\meter}$. The experimental cell with the working fluid is mounted on an electromagnetic shaker driven sinusoidally by a function generator via a power amplifier. The vertical modulation of the acceleration $a(t)=a\sin\left(2\pi f_ct\right)$ is measured by a piezoelectric accelerometer via a charge amplifier. The accelerometer is fixed to the base of the cell, allowing the measurement of the imposed modulation with a resolution of $\SI{0.1}{\meter/\square\second}$. For a given excitation frequency $f_c$, when the acceleration modulation $a$ surpasses a certain threshold $a_c$, standing waves develop on the fluid surface oscillating at $f_c/2$ through a parametric instability: the so-called Faraday waves~\cite{Faraday1831}. These standing waves interact with a set of propagating waves generated by the horizontal motion of a rectangular plunging acrylic wave maker ($6\times\SI{13}{\square\centi\meter}$) set at $\SI{1.5}{\centi\meter}$ inward from one of corners of the cell and driven by another electromechanical shaker via a power amplifier. The wave maker is excited with random noise supplied by a function generator filtered in a frequency range $2-\SI{5}{\hertz}$ by means of a band-pass filter. The output colored noise has a standard deviation $V_r\in[0.69,1.30]\,\si{\volt}$. The local wave height $\eta\left(t\right)$ is measured $\SI{5}{\centi\meter}$ away from the wave maker with a capacitive wire sensor, plunging perpendicular to the fluid surface at rest and adjoined to the experimental cell. The capacitance of the wire gauge, which changes linearly with the fluid level, is measured using a LRC circuit and a DSP Lock-in amplifier~\cite{Gordillo2011} (time constant $\SI{0.3}{\milli\second}$ and a $\SI{24}{\decibel}/$oct roll-off). The linear sensing range of the wire gauge allows measurements from $\SI{10}{\micro\meter}$ up to $\SI{30}{\milli\meter}$ with $\SI{80}{\milli\volt/\milli\meter}$ sensitivity. The data of both the fluid level and the vertical acceleration is recorded through a NI acquisition card and then analyzed in Matlab. In the experiments reported in this Letter, the acquisition time is $\SI{1}{\hour}$, the acquisition frequency is $f_{adq}=\SI{1}{\kilo\hertz}$, the excitation frequency is $f_c=\SI{20}{\hertz}$ and $a\in[1,2]\, \si{\meter/\square\second}$.

\section{Probability density function and Frequency power spectrum}

In the case where only Faraday waves are excited by vertical vibrations (once the critical acceleration to develop the parametric instability $a_c=\SI{1.08}{\meter/\square\second}$ is surpassed), $\eta(t)$ displays a discrete set of excited frequencies: the harmonics of the fundamental parametric wave frequency $f_c/2$. At $f_c=\SI{20}{\hertz}$, the typical wavelength of the standing pattern is $\SI{1.5}{\centi\meter}$. In these experimental runs, the vertical acceleration was set to $a=\SI{1.57}{\meter/\square\second}$ ($\epsilon=(a-a_c)/a_c=0.45$). In the case where only random waves are excited by the horizontal motion of the wave maker, no WT spectra (either for gravity or capillary waves) develop due to the low forcing intensity~\cite{FauveFalcon2007}. The typical wavelengths of the random waves in the frequency forcing range are between 6 to $\SI{16}{\centi\meter}$, and these waves are deep water waves~\cite{Hasselmann1962}.

\begin{figure}[h!]
\centering
\includegraphics[width=0.95\columnwidth]{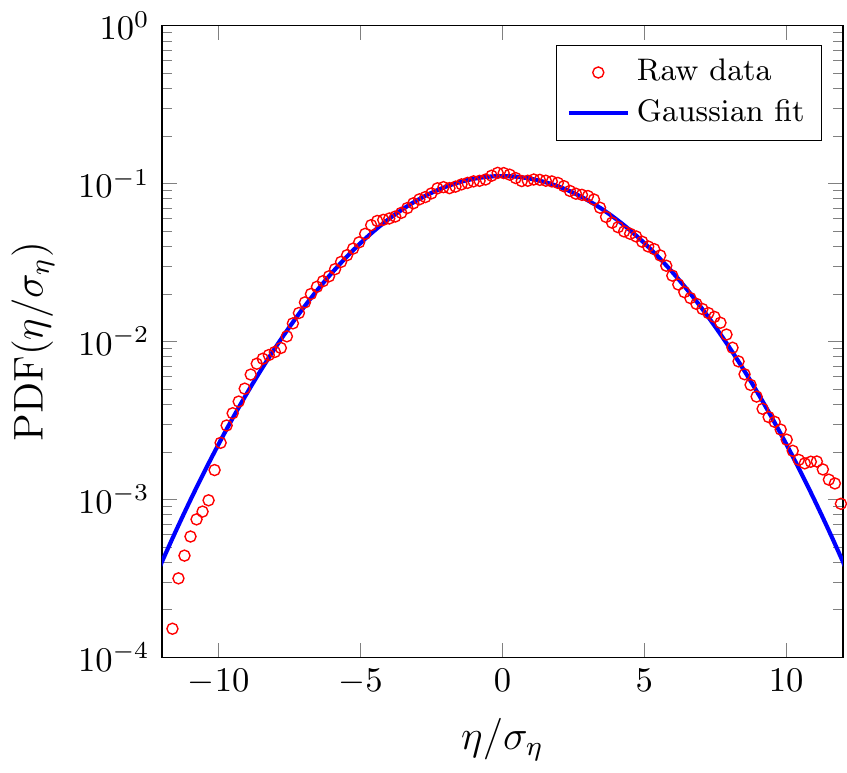}
\caption{(color online) Probability density functions of the local amplitude fluctuations height $\eta/\sigma_\eta$ for the case where random gravity waves interact with Faraday waves ({\color{red}$\circ$}). A normal curve (continuous line) is plotted for comparison.} 
\label{pdf}
\end{figure}

\begin{figure}[h!]
\centering
\includegraphics[width=0.95\columnwidth]{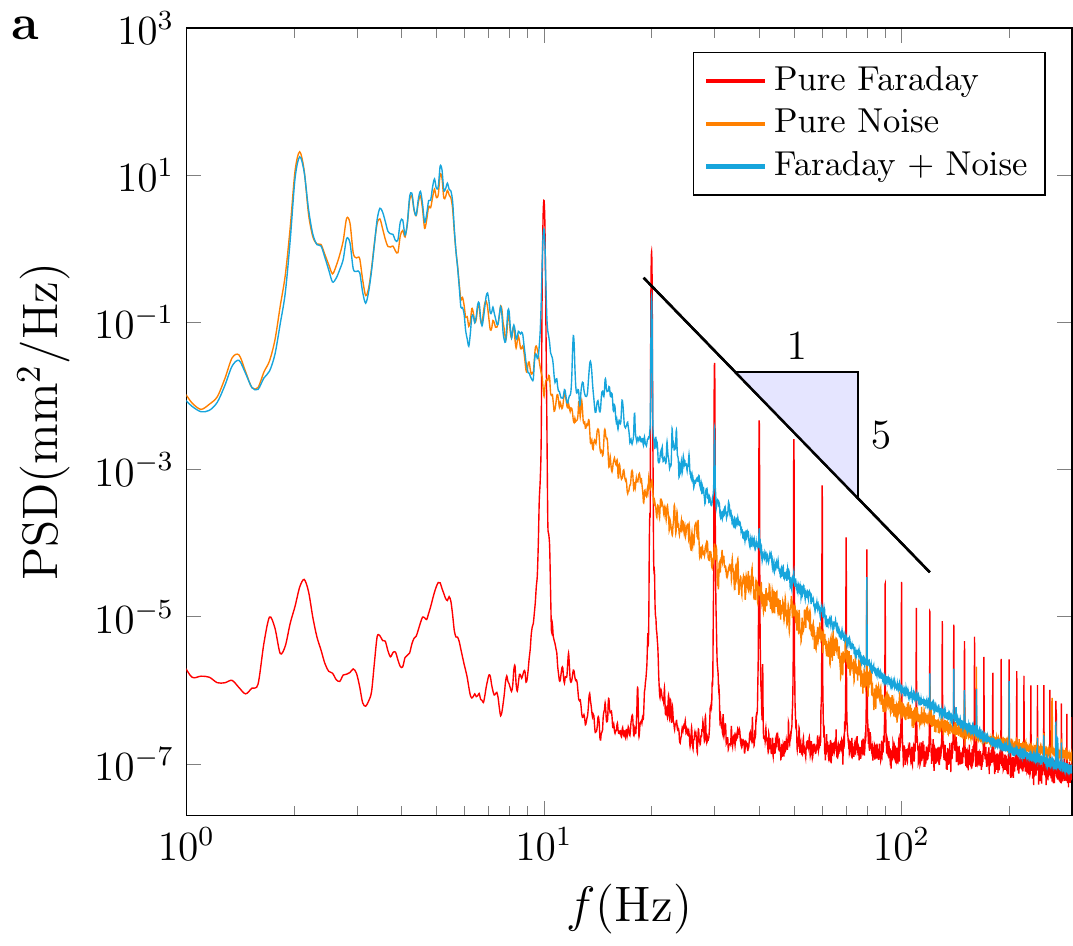}
\includegraphics[width=0.95\columnwidth]{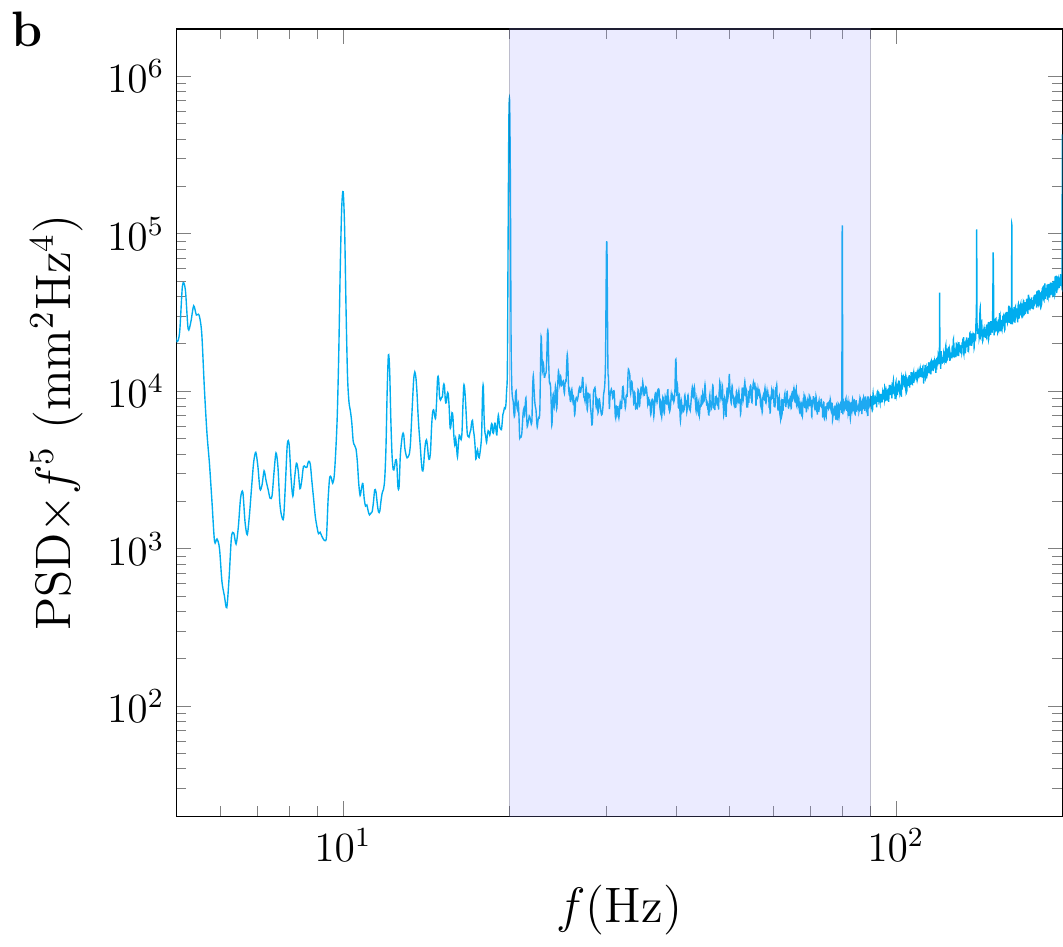}
\caption{(color online) {\bf a}: Log-log plot of the spectra $S_{\eta}(f)$ for Faraday waves, random waves, and interacting random and Faraday waves as a function of frequency $f$. The continuous line is a $f^{-5}$ best fit slope. {\bf b}: Compensated spectra $S_{\eta}(f)\times f^{5}$ for interacting random and Faraday waves.}
\label{FigEspectro}
\end{figure}

In the case where random waves are placed in interaction with Faraday waves, the probability density function (PDF) of the wave amplitude, displayed in Fig. \ref{pdf}, is Gaussian, which shows that nonlinearities in the wave amplitude are small. Besides, a scale invariant spectrum $S_{\eta}(f)\propto f^{-5}$ develops from $f_c$ to $\SI{90}{\hertz}$. This type of spectrum is observed for $\epsilon>0$ (when the parametric instability has already set in the experimental system) and for a large range of $V_r$ (in particular the one reported here for $\epsilon=0.45$). In Fig. \ref{FigEspectro}a we show $S_{\eta}(f)$ for the three configurations described above. For the last two cases, for frequencies larger than \SI{90}{\hertz}, dissipation over the wire gauge is relevant, which changes the slope of the spectra.

Two important comments related to the observation of this scale-invariant spectrum must be noticed. First, although such a large value of $\epsilon$ may sustain droplet ejection in Faraday waves~\cite{Lathrop1999} this is not observed in our experiments (either alone or coupled to randomly excited gravity waves). Thus, no wave breaking is observed. Second, experimental runs with similar values for $h$, $f_c$, $V_r$, $a_c$ were also performed on a larger cylindrical container ($\SI{30}{\centi\meter}$ in diameter), displaying the same scale-invariant spectrum. Data from these runs is not presented here because, due to large size and weight of the experimental cell, an off-axis motion appeared once the parametric waves were coupled with the random ones, which lead to a slow change in the acceleration modulation delivered by the shaker and inhomogeneities of the wave pattern. 

\begin{figure}[h]
\centering
\includegraphics[width=\columnwidth]{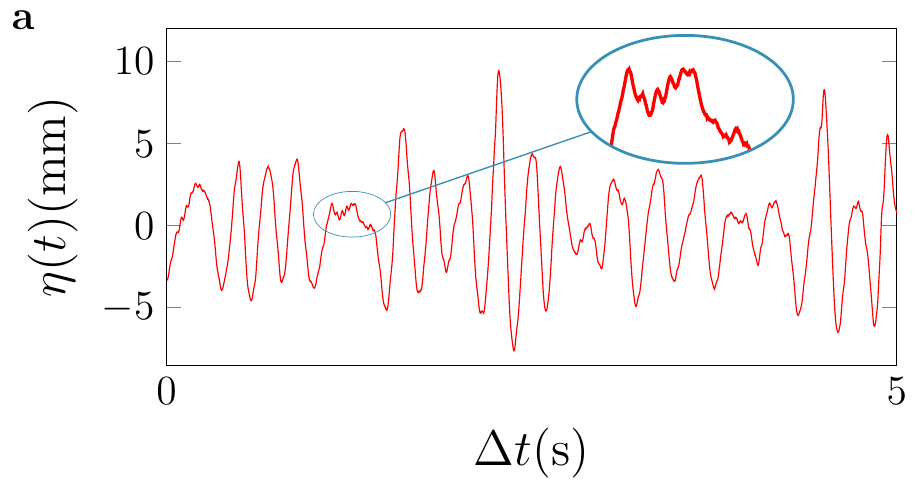}
\includegraphics[width=\columnwidth]{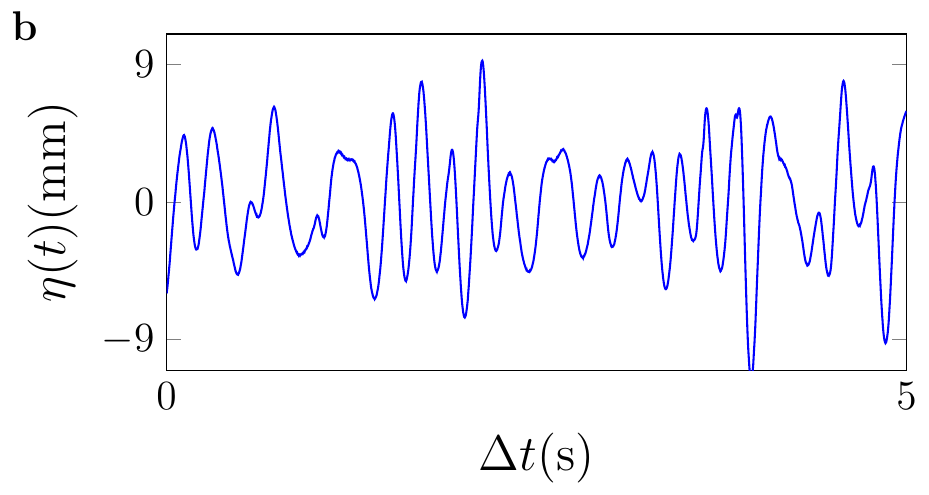}
\caption{(color online) {\bf a}: Temporal trace of the local surface wave amplitude of randomly excited waves in interaction with parametrically excited waves over a temporal window of $\Delta t = \SI{5}{\second}$ for $V_r= \SI{1.0}{\volt}$ and $\epsilon=$0.45. The zoomed region shows one of the ``anomalies" in the signal, which we believe yield the $f^{-5}$ spectrum. {\bf b}: Temporal trace of the local surface wave amplitude of randomly excited waves over a temporal window of $\Delta t = \SI{5}{\second}$ for $V_r= \SI{1.0}{\volt}$.} 
\label{FigSignal}
\end{figure}

In the case where random waves are placed in interaction with Faraday waves, the signal presents some ``anomalies", which are not displayed in the other two cases. An example of this is shown in Figure \ref{FigSignal}. It can be seen that these events do not display discontinuities in $\eta(t)$ nor in its derivative, but they do display a strongly erratic behavior. We believe the appearance of these events is responsible for the $f^{-5}$ spectrum, as it was already observed in~\cite{FauveFalcon2009}, though a thorough analysis is required to conclude on this point.

The steep spectrum might arise due to a couple of different reasons. First, fine size effects could occur. As it has been shown~\cite{Zakharov2005,Lvov2006,Nazarenko2006}, in finte size systems some quantized wavelengths cause a depletion of pure resonances, making the spectrum steeper. However, as we explained above, in our system, the $f^{-5}$ spectrum is observed in either the small container (where we can fit one half of the largest available wavelength) or the larger one (where we can fit 2 times the largest available wavelength). A second reason, might be the presence of nonlinear coherent structures, such as breaking or sharp-crested waves~\cite{Kuznetsov2004}. Depending on the spatial dimensionality of the structures, the spectrum might be $f^{-4}$ (coinciding with WT), or $f^{-5}$, the Phillips' spectrum. These structures are known to provide the main mechanism for dissipation of energy and to be related also to the phenomenon of intermittency~\cite{NazarenkoLukaschuk2010}. In our case, though, we do not observe wave breaking nor droplet ejections.
This discrepancy between theory and experiments reminds us of others works on gravity wave turbulence~\cite{Denissenko2007,Deike2012,Deike2013,Deike2015}, and on flexural wave turbulence on a metallic plate~\cite{Chaigne2001,Miquel2011,Humbert2013,Deike2014,Miquel2014}, that have shown that the occurrence of dissipation at all scales causes the energy flux to be non constant through the scales, in contrast to the WT assumptions, thus making the inertial window ill-defined, leading to a steeper spectrum. Non-local interactions might also be responsible for the steep spectrum. It is known that the interaction of a slow mode with the small scales can be interpreted as a condensate state that displays a steep spectrum~\cite{Laurie2012}. The determination of the actual reason for the development of the observed $f^{-5}$ spectrum requires further studies.

\begin{figure}[h!]
\centering
\includegraphics[width=\columnwidth]{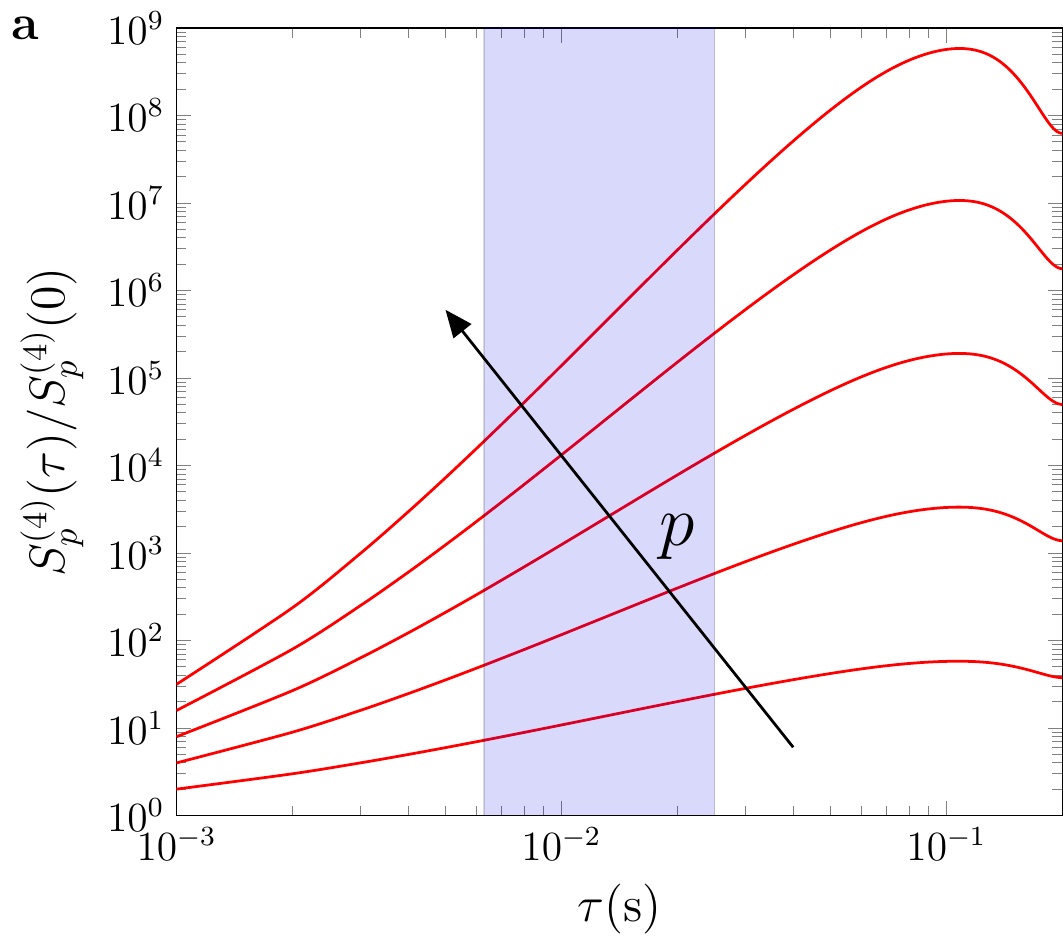}
\includegraphics[width=\columnwidth]{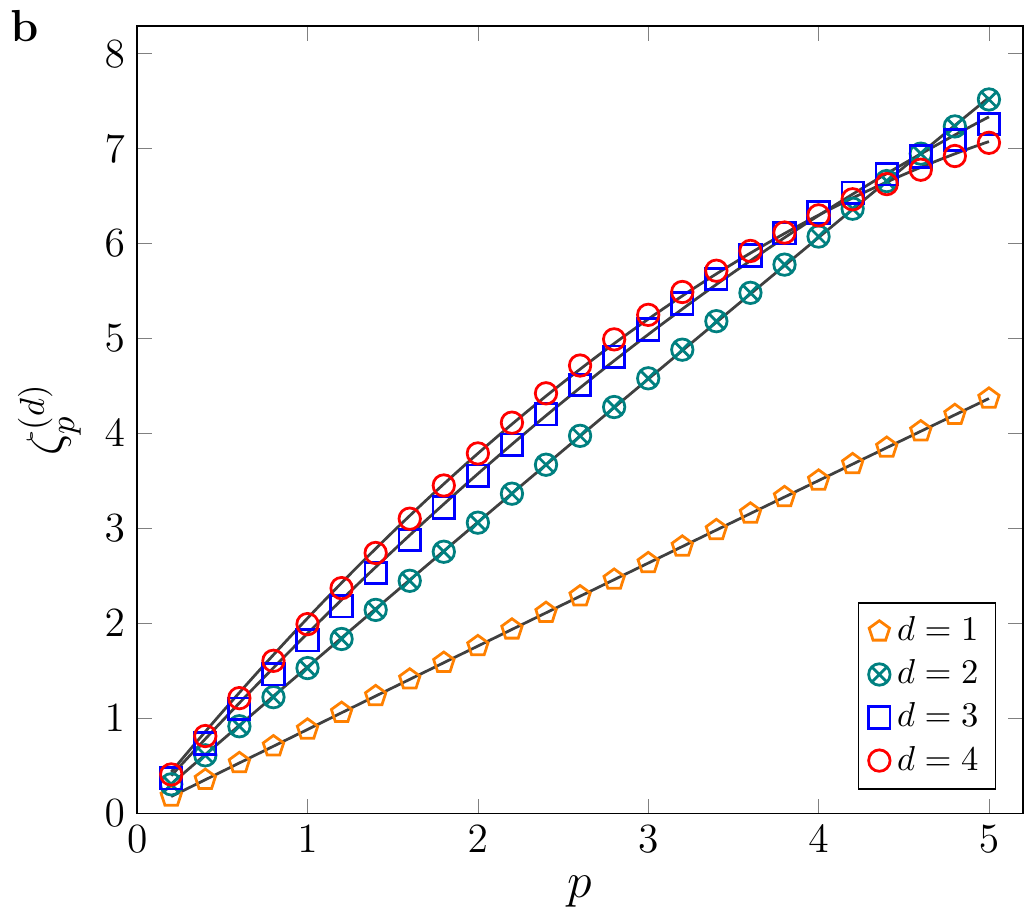}
\caption{(color online) {\bf a}: Normalized structure factors $S^{(4)}_p(\tau)/S^{(4)}_p(0)$ for $p = [1, 2, 3, 4, 5]$. They all show a power-law behavior in the inertial window $\SI{5.6}{\milli\second}<\tau<\SI{25}{\milli\second}$ (besides, the behavior is the same for different values of $d$). {\bf b}: Exponents $\zeta^{(d)}_{p}$ as a function of $p$ for $d=1,2,3$ and $4$, obtained by fitting power laws $S^{(d)}_{p}(\tau)\sim\tau^{\zeta^{(d)}_p}$. Continuous lines are best fits following the quadratic relation $\zeta^{(d)}_{p}=c_1^{(d)}p-\frac{1}{2}c_2^{(d)}p^{2}$. For $d=1$ and 2, the quadratic parameter is negligible.  Fitting $\zeta^{(4)}_{p}$ with the above relations yields $c_1^{(4)}$ = 2.21 and $c_2^{(4)} $= 0.32.}
\label{FigZetap}
\end{figure}

\section{Intermittency}
\begin{figure}[h]
\centering
\includegraphics[width=\columnwidth]{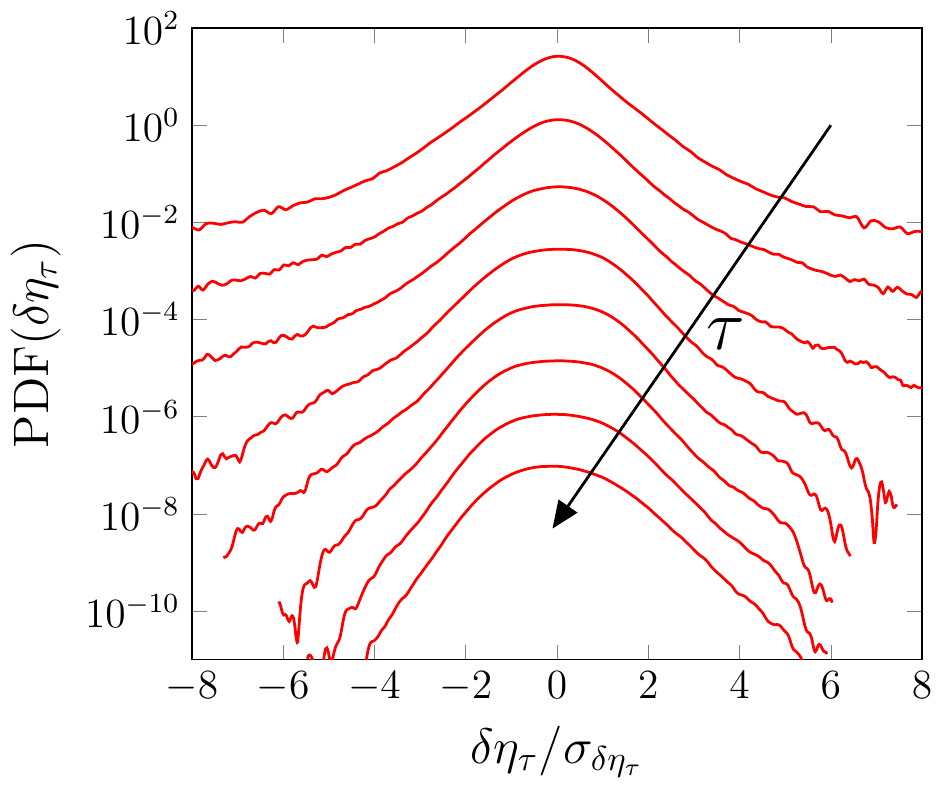}
\caption{(color online) Semilog plot of PDFs of the normalized increments $\delta^{(4)}_{\tau}\eta(t)/\langle|\delta^{(4)}_{\tau} \eta(t)|^{2}\rangle^{1/2}$ as a function of $\delta^{(4)}_{\tau}\eta(t)/\langle|\delta^{(4)}_{\tau} \eta(t)|^{2}\rangle^{1/2}$ for time lags time  $\SI{6.3}{\milli\second}<\tau<\SI{25}{\milli\second}$. Curves are displaced by a multiplying factor for better presentation. The arrow shows the direction of increasing $\tau$.}
\label{FigPDF}
\end{figure}

When a singularity spectrum develops, the statistical properties of the wave field should be affected and intermittency may appear (as in the case of the breakdown of WT). To probe this hypothesis, we calculate the higher-order cumulants of $\eta(t)$ when it displays the $f^{-5}$ spectrum. Due to the steepness of the spectrum, to probe intermittency at least third degree increments $\delta^{(3)}_{\tau}\eta(t)=\eta(t+3\tau)-3\eta(t+2\tau)+3\eta(t+\tau)-\eta(t)$ have to be computed to remove the local linear trends in the time lag $\tau$ coming from the differentiable part of $\eta(t)$~\cite{FalconRoux2010}. Accordingly, in order to study the intermittent properties of the signal $\eta(t)$, we computed the structure functions of degree $d$ and order $p$, $S^{(d)}_{p}(\tau)=\langle|\delta^{(d)}_{\tau}\eta(t)|^{p} \rangle$.  In the case of a power-law spectrum $S_{\eta}(f)\propto f^{-\nu}$ (such as the one we observed with $\nu=5$), $S^{(d)}_{p}(\tau)$ is expected to be scale as $\tau^{\zeta^{(d)}_p}$, and in particular $S^{(d)}_{2}(\tau)\sim\tau^{\nu-1}$. In Figure \ref{FigZetap}a we show the behavior of $S^{(4)}_{p}(\tau)=\langle|\delta^{(4)}_{\tau}\eta(t)|^{p} \rangle$. It is observed that the structure factors are indeed power-laws as a function of $\tau$ (for any degree $d$). Thus, we can compute the exponents $\zeta^{(d)}_p$ as a function of $p$ for $d=1,2,3$ and 4 by best fitting a power law in the interval $\SI{5.6}{\milli\second}<\tau<\SI{25}{\milli\second}$ (the inertial window). These exponents are shown in Fig. \ref{FigZetap}b.  It can be observed that for $d=1$ and 2, a linear relationship is obtained between the exponent and $p$, which shows the differentiable part of the $\eta(t)$. However, when we look at the higher-degree exponents we realize that these exponents are non-linear functions of $p$, which is a clear signature of the intermittent nature of the wave system. Moreover, these exponents are in good agreement with the theoretical prediction $\zeta^{(d)}_{p}=c_1^{(d)}p-\frac{1}{2}c_2^{(d)}p^{2}$~\cite{FalconRoux2010}. Additionally, we note that whereas using $d=1$ or 2 the exponents lead to an underestimated value of $\nu=5$, the exponents $\zeta^{(3)}_{2}$=3.58 and $\zeta^{(4)}_{2}$ = 3.79, which show that by increasing $d$ one approaches the observed spectral exponent. Another feature of the intermittent nature of the acquired signal can be observed by computing the PDFs of the normalized increments $\delta^{(d)}_{\tau}\eta(t)$ for different values of $\tau$. In our case, since we concluded above that we should use $d=4$, we computed $\delta^{(4)}_{\tau}\eta(t)$ for different time lags $\tau$ and with them we calculated their PDFs, which are shown in Fig. \ref{FigPDF}. We note that their shapes deform continuously as the time lag $\tau$ changes, showing an evolution across scales. Within the inertial window, for small $\tau$ the tails of the PDF show a very mild slope which decreases for larger values of $\delta^{(4)}_{\tau}\eta(t)$, while for larger $\tau$ the tails become much steeper. This effect reaffirms the previous results on the intermittent nature of the wave field, which shows larger fluctuations for smaller time lags ({\it i.e.} small scale intermittency). Thus, the observed wave field displays a scale-invariant power-law spectrum between 20 and $\SI{90}{\hertz}$ and it is intermittent at small scales.

The origin of this intermittent wave field lies in the interaction between the random and parametric waves. The randomly moving piston generates gravity waves which distort the pattern of Faraday waves, changing locally its amplitude and wavelength, and even creating defects. These distortions pass over the wire gauge, oscillating at a much larger frequency than the random wave that act as low frequency carriers. In that sense, it follows the same idea developed by Phillips~\cite{Phillips1958}, but without the need of wave breaking. In fact, no wave breaking nor droplet ejection was observed in the experimental runs. It must be noticed that a similar spectrum can be constructed by following the amplitude defects of Faraday waves which are self-generated as $a$ is increased above the threshold of defect turbulence~\cite{FauveFalcon2009}. This suggests a connection between the observed spectrum and the spectrum of wave amplitudes in the defect-mediated dynamics of pattern forming systems. In those systems, non locality is needed to generate turbulent states. In our experimental configuration it is observed that the control parameter range for the appearance of this scale-invariant spectrum is very large and insensitive to the values of $f_c$, $a_c$ (as long as is large enough to sustain Faraday waves over the random ones) and $V_r$, which shows the robustness of this type of spectrum.

\section{Conclusion}
In conclusion, we have shown that a simple tabletop experiment can display a power-law spectrum driven by singularities when a standing wave pattern arising from a parametric instability is coupled with randomly excited gravity waves. The wave field generated by this coupling is intermittent as it can be observed by the nonlinear dependence of the $\zeta^{(d)}_p$ exponents on the order $p$, and by the change in the shape of the PDF of the normalized increments $\delta^{(d)}_{\tau}\eta(t)$ when $\tau$ changes. Some questions related to the wave field spectrum remain unanswered (such as the possible multifractal structure of the intermittent increments and its structural relation with the breakdown of WT spectrum, or the mechanisms involved behind the steep spectrum). To answer some of them, spatio-temporal measurements will be needed. Work is being performed in that direction. 

\section{ACKNOWLEDGMENTS}
This research was supported by Fondecyt Grant No. 3160032 (G.C).


\begin{thebibliography}{99}

\bibitem{KolmogorovWTBook}  V. E. Zakharov, V. S. L'vov, and G. Falkovich, {\it Kolmogorov Spectra of Turbulence I: Wave Turbulence} (Springer Series in Nonlinear Dynamics, Springer, 1992); S. Nazarenko, {\it Wave Turbulence} (Lecture Notes in Physics, Springer, 2011); A. C. Newell and B. Rumpf, Annu. Rev. Fluid Mech. {\bf43} 59-78 (2011).

\bibitem{Hasselmann1962} K. Hasselmann, J. Fluid Mech. {\bf12}, 481 (1962); {\bf15}, 273 (1963).

\bibitem{ZakharovDiachenko2004} M. Onorato {\it et al.}. Phys. Rev. Lett. {\bf89}, 144501 (2002); A. Pushkarev, D. Resio, and V. Zakharov, Physica
(Amsterdam) {\bf184D}, 29 (2003); A. I. Dyachenko, A. O. Korotkevich, and V. E. Zakharov, Phys. Rev. Lett. {\bf92}, 134501 (2004)

\bibitem{Toba1973} Y. Toba, J. Oceanogr. Soc. Jpn. {\bf29}, 209 (1973);  K. K.
Kahma, J. Phys. Oceanogr. {\bf11}, 1503 (1981).

\bibitem{FauveFalcon2007} E. Falcon, C. Laroche, and S. Fauve
Phys. Rev. Lett. {\bf98}, 094503 (2007).

\bibitem{ZakharovFilonenko1967} V. E. Zakharov and N. N. Filonenko, J. Appl. Mech. Tech. Phys. {\bf4}, 506Ð515 (1967).

\bibitem{PushkarevZakharov1995} A. N. Pushkarev and V. E. Zakharov, Phys. Rev. Lett. {\bf76}, 3320 (1996); L. Deike, D. Fuster, M. Berhanu, and E. Falcon Phys. Rev. Lett. {\bf 112}, 234501 (2014).  

\bibitem{Putterman1996} W. B. Wright, R. Budakian, and S. J. Putterman, Phys. Rev. Lett. {\bf76}, 4528 (1996); M. Lommer and M. T. Levinsen, J. Fluoresc. {\bf12}, 45 (2002); E. Henry, P. Alstr\o m, and M. T. Levinsen, Europhys. Lett. {\bf52}, 27 (2000); C Falcon, E Falcon, U Bortolozzo, S Fauve, Europhys. Lett {\bf86} (1), 14002 (2009). 

\bibitem{DuringRica2006} G. D\"uring, C. Josserand, and S.Rica
Phys. Rev. Lett. {\bf97}, 025503 (2006). 

\bibitem{Mordant2008} N. Mordant
Phys. Rev. Lett. 100, 234505 (2008); A. Boudaoud, O. Cadot, B.Odille, and C. TouzŽ, Phys. Rev. Lett. {\bf100}, 234504 (2008).

\bibitem{Dyachenko1992} S. Dyachenko, A. Newell, A. Pushkarev, and V. Zakharov, Physica D {\bf57}, 96 (1992), C. Connaughton {\it et al},  Phys. Rev. Lett. {\bf95} 236901 (2005).

\bibitem{Bortolozzo2009} U. Bortolozzo, J. Laurie, S. Nazarenko, and S. Residori, J. Opt. Soc. Am B {\bf26} (12) 2280Ð2284 (2009).

\bibitem{Biven2001} L. Biven, S.V. Nazarenko, and A.C. Newell, Physics Letters A {\bf280} 28Ð32 (2001).

\bibitem{During2009} G. DŸ\"uring, A. Picozzi, and S. Rica, Physica D {\bf238}, 1524Ð1549  (2009).

\bibitem{Phillips1958} O. M. Phillips, J. Fluid Mech. {\bf4}, pp. 426-434 (1958).

\bibitem{Mitsuyasu1974} H. Mitsuyasu, and T. Honda, J. Oceanogr. Soc. Jpn. {\bf30},185-198 (1974).

\bibitem{Romero2012} L. Romero,  W. K. Melville, and J. M. Kleiss, J. Phys. Oceanogr. {\bf42}, 1421Ð1444 (2012)

\bibitem{NazarenkoLukaschuk2010} S. Nazarenko, S. Lukaschuk,  S. McLelland, and  P. Denissenko, J. Fluid Mech. {\bf642}, 395Ð420 (2010).

\bibitem{NewellZakharov2008} A. C. Newell, V. E. Zakharov, Phys. Lett. A {\bf372}, 4230Ð4233 (2008).

\bibitem{NewellZakharov1992} A. C. Newell, V. E. Zakharov,  Phys. Rev. Lett. {\bf69}, 1149 (1992)

\bibitem{Miquel2013} B.Miquel, A. Alexakis, C. Josserand, and N: Mordant, Phys. Rev. Lett. {\bf111}, 054302 (2013), S. Chibaro and C. Josserand, Phys. Rev. E {\bf94}, 011101(R).

\bibitem{Kuznetsov2004} E. A. Kuznetsov, JETP Lett. {\bf80}, (2), 83Ð89 (2004).

\bibitem{Laurie2010} J. Laurie, V. S. L'vov, S. Nazarenko, and O. Rudenko, Phys. Rev. B {\bf 81}, 104526 (2010).

\bibitem{Lvov2010} V. S. L'vov, and S. Nazarenko, JETP (Pisma Zh. Eksp. Teor. Fiz) {\bf 91} 464-470 (2010).

\bibitem{Balk1990} A. M. Balk, S. V. Nazarenko, and V. E. Zakharov, Physics Letters A, {\bf146}(4), 217-221 (1990).

\bibitem{Nazarenko1991} S. V. Nazarenko,  Pisma Zh. Eksp. Teor. Fiz {\bf53}(12), 604-607 (1991).

\bibitem{Connaughton2011} C. Connaughton, S. Nazarenko, and B. Quinn, EPL, {\bf96}(2), 25001?6 (2011).

\bibitem{Mehtar2017} Y. Mehtar-Tani, Nucl. Phys. A 966, 241 (2017).

\bibitem{Faraday1831} M. Faraday, Phil. Trans. R. Soc. London {\bf 121}, 299 (1831).

\bibitem{Gordillo2011} L. Gordillo {\it et al}, Eur. Phys. J. D {\bf62}, 39Ð49 (2011).

\bibitem{Lathrop1999} C. L. Goodridge, H. G. E. Hentschel, and D. P. Lathrop, Phys. Rev. Lett. {\bf82}, 3062 (1999).

\bibitem{FauveFalcon2009} C. Falc\'on and S. Fauve, Int. J. Bifurcat. Chaos {\bf19}, 3553 (2009).

\bibitem{Zakharov2005} V. E. Zakharov {\it et al.} JETP {\bf82}(8), 487-491 (2005).

\bibitem{Lvov2006} Y. V. Lvov, S. Nazarenko, and B. Pokorni, Physica D {\bf 218}(1), 24-35 (2006).

\bibitem{Nazarenko2006} S. Nazarenko, J. Stat. Mech. L02002 (2006)

\bibitem{Denissenko2007} P. Denissenko, S. Lukaschuk,  and S. Nazarenko, Phys. Rev. Lett. {\bf 99}, 014501 (2007).

\bibitem{Deike2012} L. Deike, M. Berhanu, and E. Falcon, Phys. Rev. E, {\bf 85}(6), 066311 (2012).

\bibitem{Deike2013} L. Deike, J.-C. Bacri, and E. Falcon, J. Fluid Mech. {\bf733}, 394-413 (2013).

\bibitem{Deike2015} L. Deike {\it et al}, J. Fluid Mech. {\bf781}, 196?225 (2015).

\bibitem{Chaigne2001} A. Chaigne and C. Lambourg, JASA {\bf109}, 1422-1432 (2001).

\bibitem{Miquel2011} B. Miquel, and N. Mordant, N. Phys. Rev. Lett. {\bf107}, 034501 (2011).

\bibitem{Humbert2013} T. Humbert {\it et al.} EPL {\bf102}, 30002 (2013).

\bibitem{Deike2014} L. Deike, M. Berhanu, and E. Falcon, Phys. Rev. E {\bf89}, 023003 (2014).

\bibitem{Miquel2014} B. Miquel, A. Alexakis, and N. Mordant, Phys.Rev.E {\bf89}, 062925?10 (2014).

\bibitem{Laurie2012} J. Laurie, U. Bortolozzo, S. Nazarenko, and S. Residori, Phys. Rep. {\bf514}, 121175 (2012).

\bibitem{FalconRoux2010} E. Falcon, S. G. Roux, and B. Audit, Europhys. Lett {\bf90}, 50007 (2010).


\end{thebibliography}
\end{document}